\documentclass[conference]{IEEEtran}

\usepackage{graphicx}

\usepackage{times}

\usepackage{amsfonts}
\usepackage{amsmath,amssymb}
\usepackage{ams}

\begin{document}

% paper title
\title{Capabilities Engineering:\\ Constructing Change-Tolerant Systems}

\author{\authorblockN{Ramya Ravichandar, James D. Arthur, Shawn A. Bohner}
\authorblockA{Department of Computer Science\\
Virginia Polytechnic Institute and State University\\
Blacksburg, Virginia 24060\\
Email: \{ramyar, arthur, sbohner\}@vt.edu}}

\maketitle

%  EXACTLY 149 words
\begin{abstract}
We propose a Capabilities-based approach for building long-lived, complex systems that have lengthy development cycles. User needs and technology evolve during these extended development periods, and thereby, inhibit a fixed requirements-oriented solution specification. In effect, for complex emergent systems, the traditional approach of baselining requirements results in an unsatisfactory system. Therefore, we present an alternative approach, \textit{Capabilities Engineering}, which mathematically exploits the structural semantics of the Function Decomposition graph --- a representation of user needs --- to formulate Capabilities. For any given software system, the \textit{set} of derived Capabilities embodies change-tolerant characteristics. More specifically, each individual Capability is a functional abstraction constructed to be highly cohesive and to be minimally coupled with its neighbors. Moreover, the Capability set is chosen to accommodate an incremental development approach, and to reflect the constraints of technology feasibility and implementation schedules. We discuss our validation activities to empirically prove that the Capabilities-based approach results in change-tolerant systems.
\end{abstract}

\IEEEpeerreviewmaketitle

\section{Introduction}
In the more recent times there has been an increase in the development of large-scale complex systems. Hybrid communication systems, state-of-art defense systems, technologically advanced aeronautics systems and similar engineering projects demand huge investments of time and money. Unfortunately, a number of such systems fail or are prematurely abandoned despite the availability of necessary resources \cite{Glass1998} \cite{Deal2004} \cite{Jazequel1997}. 
A major reason for this is that they have lengthy development periods during which various factors of change detrimentally influence the system. The effect of factors such as changing user needs, varying technology constraints and unpredictable market demands is further exacerbated by the inherent complexity of these systems. The primary manifestation of change, irrespective of its cause, is in the form of requirements volatility; requirements are specifications that dictate system development.
It is recognized that the impact of requirements volatility has far-reaching effects like increasing the defect density during the coding and testing phase \cite{Malaiya1999} and affecting the overall project performance \cite{Zowghi2002}. 

Traditional Requirements Engineering (RE) attempts to minimize change by baselining requirements. This approach is successful in small-scale systems whose relative simplicity of system functionality and brief development cycles discourages changing user perceptions. Furthermore, the inability to foster new technology in a short time-period assures the realization of systems using initial technology specifications. In contrast, traditional RE is ill-equipped to scale the monumental complexity of large-scale systems and accommodate the dynamics of extended development periods. Hence, there is a need for an alternative approach that transcends the complexity and scalability limits of current RE methods. 
  
We propose a Capabilities-based approach for building large-scale complex systems that have lengthy development cycles. Capabilities are \textit{change-tolerant functional abstractions} that are foundational to the composition of system functionality. User needs are the primary sources of information about desired system functions. We use a \textit{Function Decomposition} (FD) graph to represent needs, and thereby, understand desired system functionalities and their associated levels of abstraction. The \textit{ Capabilities Engineering} (CE) process mathematically exploits the structural semantics of the FD graph to formulate Capabilities as functional abstractions with high cohesion and low coupling. The process also employs a multi-disciplinary optimization (MDO) approach to select optimal sets of Capabilities that accommodate an incremental development approach, and  reflect the constraints of technology feasibility and implementation schedules.
Note that Capabilities are architected to accommodate specific factors of change, \textit{viz.} requirements volatility and technology advancement. 
We conjecture that the impact of requirements volatility is less likely to propagate beyond the affected Capability because of its reduced coupling with neighboring Capabilities. Additionally, the property of  high cohesion helps localize the impact of change to within a Capability. The other factor of change, technology advancement, is accounted for by the conscious assessment of technology feasibility as a part of the MDO approach. Therefore, Capabilities are intentionally constructed to possess characteristics that accommodate the major factors of change. In fact, we envision CE as a possible solution to the research challenge of evolving Ultra-Large-Scale systems \cite{Northrop2006}.

The remainder of the paper is organized as follows: Section II discusses characteristics of both large-scale and conventional systems, examines general strategies for managing change, and provides a review of related work. Section III  outlines the overall process of CE, and details the formulation and optimization of Capabilities. Also, metrics to measure coupling and cohesion of Capabilities are defined. Section IV outlines the validation activities and discusses preliminary observations. Our conclusions are given in Section V.

\section{Change Management Strategies}

We define \textit{complex emergent systems} as systems that are large-scale, complex, have lengthy development cycles and have a lifetime of several decades. A system is said to be complex when it consists of a large number of parts that interact in a non-trivial manner \cite{Simon1969}. The colossal magnitude of a large-scale complex system impedes \textit{a priori} knowledge about the effects of these interactions. 
As a result, the behavioral characteristics of the overall system is greater than a mere aggregation of its constituent elements. This behavior includes properties that \textit{emerge} from the elemental interactions and are characteristic only of the global system. Specifically, it is fallacious to attribute these emergent properties to individual elements of the system \cite{Heylighen1989}. Unlike complex emergent systems, conventional systems are smaller-scale, less complex, have brief development cycles and have a shorter lifetime. Consequently, requirements can be baselined after a certain point in the development period. However, requirements and technology often evolve during the extended development periods of complex emergent systems, and thereby, inhibit a comprehensive up-front solution specification. Thus, a primary difference between developing conventional software systems and complex emergent systems is the lack of a final solution specification in the latter case caused by continuous system evolution.

The first law of software evolution \cite{Lehman1996}, asserts that if a system is to be satisfactory it has to constantly adapt to change. One can pursue either of the two following strategies to reconcile with change:

\subsection{Minimize Change}
\label{sec:RequirementsEngineering}
Traditional RE attempts to \textit{minimize change} by baselining requirements prior to design and development. This mandates that needs, the originating source of requirements, be accurate and complete. Different elicitation techniques such as interviews, questionnaires, focus groups, and, introspection are employed to derive needs effectively \cite{Goguen1993}. Also, numerous models strive to combat requirements volatility through iterative processes and incremental development \cite{Boehm1986} \cite{Arthur2005} \cite{Graham1992}. Some development paradigms, like Extreme Programming \cite{Beck1999}, adopt an unconventional approach of eliminating formal RE from their process. Agile Modeling proposes lightweight RE \cite{Paetsch2003}. Nevertheless, neither has been proven to work, repeatedly, for large and complex projects \cite{Keefer2002}. When building large systems, empirical research evidence  indicates the failure of traditional RE to cope with the attendant requirements evolution \cite{Bell1976} \cite{Lutz1993}. Consequently, in the case of complex emergent systems, which are often mission-critical, such failures are extremely expensive in terms of cost, time and human life. 

\subsection{Accommodate Change}
\label{sec:AccommodateChange}
Performance based specifications \cite{Performance1999} \cite{Performance2000} were introduced with the objective of \textit{accommodating} instead of minimizing change. These specifications
are statements of requirements described as  outcomes desired of a system from a high level perspective. As a result, the solution is constrained to a much lesser degree and provides greater latitude in incorporating suitable design techniques and technology. More recently, Capability Based Acquisition (CBA) \cite{Montroll2003} \cite{Biggs2003} is being used to resolve problems posed by lengthy development periods and increased system complexity. It is expected to accommodate change and produce systems with relevant capability and current technology by delaying requirement specifications in the software development cycle, and maturing a promising technology before it becomes a part of the program. 

However, neither Performance based specification nor the CBA approach defines the level of abstraction at which a specification or Capability is to be described. Furthermore, they neglect to outline any scientific procedure for deriving these types of specifications from the initial set of user needs. Therefore, these approaches propose solutions that are not definitive, comprehensive or mature enough to accommodate change and benefit the development process of complex emergent systems. Nevertheless, they do provide certain key concepts ---
reduced emphasis on detailed requirements specification, and nurturing a promising technology before it becomes a part of the program --- that are incorporated in CE as a part of its strategy to accommodate change. 

Similar to Performance based specifications and CBA, the CE process utilizes a specification-centric approach to accommodate change. It enumerates Capabilities desired of the system at various levels of abstraction. Capabilities are identified after needs analysis but prior to requirements, and indicate the functionality desired of a system at various levels of abstraction. This approach complements the research by Hevner \textit{et al.} \cite{Hevner2005}, which focuses on automatically determining the functionality of complex information systems from requirement specifications, design or program implementations. 
Capabilities are formulated to embody high cohesion and minimal coupling, and are subjected to an MDO approach to induce desirable design characteristic. Embedding desirable design traits in a specification, introduces aspects of a process-centric approach. Hence, we theorize that CE is a hybrid approach of both the process and specification-centric approach to accommodating change. 

\section{Capabilities Engineering Process}

\begin{figure}[htbp]
\centering
\includegraphics[trim = 8mm 180mm 0mm 13mm, clip, width=10 cm]{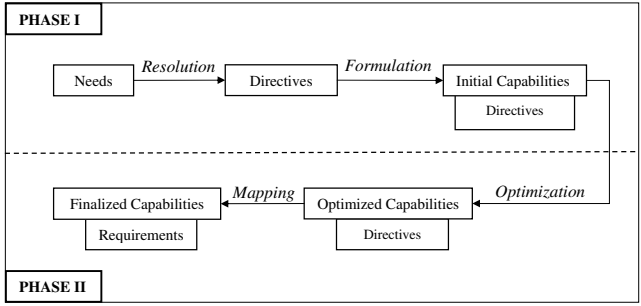}
\caption{\em Capabilities Engineering Process}
\label{fig:FigNeedsToRequirements}
\end{figure}

The CE process architects Capabilities as highly cohesive, minimally coupled functional abstractions that accommodate the constraints of technology feasibility and implementation schedule. Figure \ref{fig:FigNeedsToRequirements} illustrates the two major phases of this process. 
%------------------------------------------------------------------------ 		
%	FIGURE: Capabilities Engineering Phase
%--------------------------------------------------------------------------
%---------------------------------------------------------------------------- 
%END FIGURE: Capabilities Engineering Phase
%-----------------------------------------------------------------------------
Phase I establishes initial sets of Capabilities based on their values of cohesion and coupling. These measures are mathematically computed from an FD graph, which represents the user needs and \textit{directives}. Directives are system characteristics resolved from user needs and assist in the formulation of Capabilities. Hence, the two major activities of this phase are \textit{resolving} directives from needs and \textit{formulating} Capabilities using the FD graph. 

Phase II, a part of our current ongoing research, employs an MDO approach on the initial sets of  Capabilities to determine sets that are optimal with respect to the constraints of technology and schedule. These optimal Capabilities are then mapped to requirements as dictated by an incremental development process. Thus, the final set of Capabilities and their associated requirements constitutes the output of the CE process. Therefore, the major activities of Phase II are the \textit{optimization} of initial Capabilities and the \textit{mapping} of optimized Capabilities to requirements. 

The following sections discuss these phases and their activities in detail.

\subsection{Phase I: Resolution}
Resolution is the process of deriving directives from needs using the FD graph. First, we explain the concept of directives and the purpose of introducing them. Then, we define the elements of an FD graph and enumerate the rules for its construction. In the process, the activity of resolution is described.

\subsubsection{Directives}
Needs are elicited using various techniques \cite{Goguen1993} from different user classes to get a complete perspective of the system to be built. However, these needs may be vague, inconsistent and conflicting. Therefore, we introduce the concept of \textit{directives} to refine and resolve needs, and express system characteristics in a more consistent format. We define a directive as a characteristic formulated in the problem domain language, but described from the system's perspective. A directive has three main purposes:

\begin{itemize}
	\item \textit{Captures domain information:}
A directive can be incomplete, unverifiable, and untestable. However, it serves the purpose of describing system functionality in the language of the problem domain, which aids in capturing domain information. In contrast, a requirement that is a statement formulated in the technical language of the solution neglects to preserve and convey valuable domain information. In fact, Zave and Jackson \cite{Zave1996} have identified the lack of appropriate domain knowledge in the process of requirements refinement as a key problem area in RE. Therefore, the introduction of directives provides %additional
momentum in bridging the gap between needs and requirements.  

\item \textit{Facilitates formulation of Capabilities:}
Initial sets of Capabilities are functional abstractions that have high cohesion and low coupling. In order to formulate these initial Capabilities we need to examine all possible functional abstractions of a system. Although, directives are characteristics in the problem domain, they are implicitly associated with some functionality desired of the actual system. Hence, each functional abstraction is linked with a set of directives. In other words, a Capability is associated with a specific set of directives. Therefore, directives can be used to determine the cohesion and coupling values of potential functional abstractions, and thus assist in the formulation of Capabilities.

\item \textit{Maps to requirements:}
A directive is affiliated with the problem domain and a requirement with the solution domain; yet both share the same objective of describing the characteristics expected of the desired system. In addition, they are described at a similar level of abstraction. Hence, we conjecture that the mapping of a directive to a requirement is straightforward. As Capabilities are already associated with a set of directives, this mapping process produces requirements that form the output of the CE process. Thus, directives assist in the final activity (see Figure \ref{fig:FigNeedsToRequirements}) of mapping Capabilities to requirements.
\end{itemize}

\subsubsection{Function Decomposition Graph}
\begin{figure}[htbp]
\centering
% trim = left bottom right top
\includegraphics[trim = 10mm 170mm 0mm 15mm, clip, width=9.5 cm]{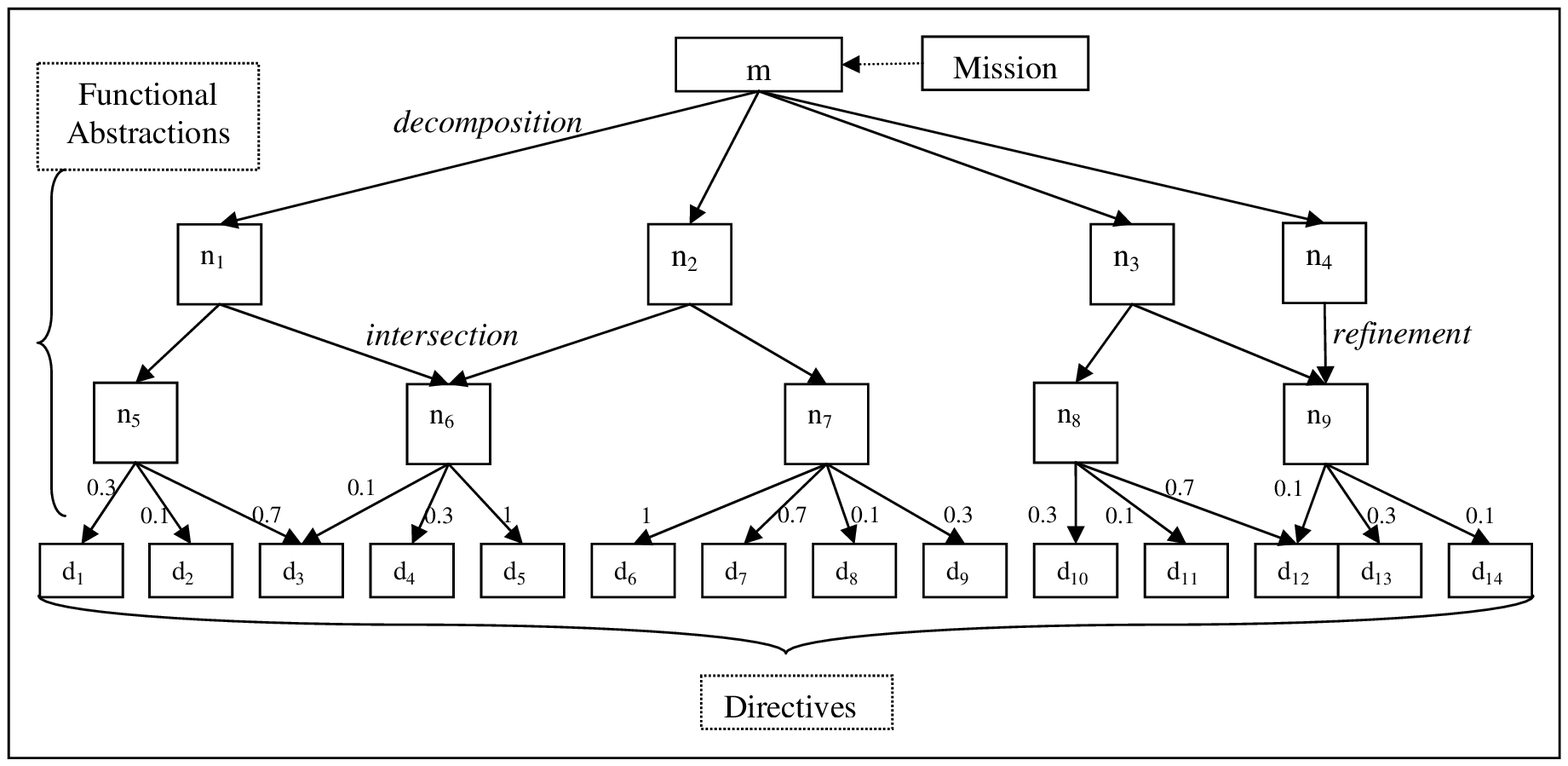}
\caption{\em Example FD Graph $G=(V,E)$}
\label{Fig-Example-Metric}
\end{figure}

Needs are the basis for understanding the functionality desired of a system. Often, needs are expressed by users at varying levels of abstraction. An abstraction presents information essential to a particular purpose, ignoring irrelevant details. In particular, a functional abstraction indicates the functionality expected of the system from a high-level perspective while ignoring minute details.
We use an FD graph to represent functional abstractions of the system, which are obtained by the systematic decomposition of user needs (see Figure \ref{Fig-Example-Metric}). A need at the highest level of abstraction is the mission of the system. This can be decomposed into other needs. We say that a decomposition of needs is equivalent to a decomposition of functions because a need essentially represents some functionality of the system. Hence, decomposition is an intuitive process of recursively partitioning a problem until an atomic level (here a directive) is reached. Specifically, the FD graph illustrates user needs in terms of desired functionalities and captures their associated levels of abstraction. In addition, the structure of the graph is reflective of the dependencies between these functionalities. Formally, we define an FD graph $G=(V,E)$ as an acyclic directed graph where: 
\begin{itemize}
	\item $V$ is the vertex set that represents system functionality at various levels of abstraction in accordance to the following rules:
	
\begin{itemize}
	\item \textbf{Mission:} The root represents the \textit{highest level mission or need} of the system.  There is exactly one overall system mission and hence, only one root node in an FD graph. In Figure \ref{Fig-Example-Metric}, $m$ is the root node as $indegree(m)=0$ (indegree is the number of edges coming into a vertex in a directed graph).
		
	\item \textbf{Directive:} The leaf node represents a \textit{directive} of the system. A system has a finite number of directives and hence, its FD graph also has the same number of leaves. In Figure \ref{Fig-Example-Metric}, nodes $d_i, i=1 \ldots 14$ represent directives as $outdegree(d_i)=0$ (outdegree is the number of edges going out of a vertex in a directed graph). 
				
	\item \textbf{Functional Abstraction:} An internal node represents a functionality of the system. The level of abstraction of the functionality is inversely proportional to the length of the directed path from the root to the internal node representing the concerned functionality. In Figure \ref{Fig-Example-Metric}, nodes $n_i, i=1 \ldots 9$ represent functional abstractions as $outdegree(n_i) \neq 0$ and $indegree(n_i) \neq 0$. 
\end{itemize}
			
\item  $E = \{(u,v) | u, v \in V, u \neq v \}$ is the edge set, where each edge indicates decomposition, intersection or refinement relationship between nodes. The edge construction rules are described below: 
\begin{itemize}
\item \textbf{Decomposition:} The partitioning of a functionality into its constituent components is depicted by the construction of a decomposition edge.  The direct edge between a parent and its child node represents functional decomposition and implies that the functionality of the child is a proper subset of the parent's functionality. For example in Figure \ref{Fig-Example-Metric} the edges $(m,n_1), (m,n_2), (m,n_3), (m,n_4)$ indicate that the functionality of $m$ is decomposed into smaller functionalities $n_1, n_2, n_3, n_4$ such that $m \equiv n_1 \underset{\scriptstyle fn} \cup n_2 \underset{\scriptstyle fn} \cup n_3 \underset{\scriptstyle fn} \cup n_4$	where $\underset{\scriptstyle fn} \cup$ is the union operation performed on functionality. Hence, only non-leaf nodes \textit{i.e.} internal nodes with an outdegree of at least two can have valid decomposition edges with their children. 
					
				\item \textbf{Refinement:} The refinement relationship is used when there is a need to express a node's functionality with more clarity, say, by furnishing additional details.	If $outdegree(u)=1, u \in V$ and $(u,v) \in E$ then the edge $(u,v)$ represents a refinement relationship. $v$ is a refined version of its parent $u$. In Figure \ref{Fig-Example-Metric}, nodes $n_4$ and $n_9$ share a refinement relationship. 
				
				\item \textbf{Intersection:} To indicate the commonalities between functions defined at the same level of abstraction the intersection edge is used. Hence, a child node with an indegree greater than one represents a functionality common to all its parent nodes. For example, in Figure \ref{Fig-Example-Metric} $n_6$ is a child node of parent nodes $n_1$ and $n_2$. Consequently, 
$n_6 \equiv n_1 \underset{\scriptstyle func} \cap n_2 $ where $ \underset {\scriptstyle func} \cap$ is the intersection operation performed on functionality. The edges $(n_1, n_6), (n_2, n_6)$ represent the intersection relationship.
						
\end{itemize}
\end{itemize}

%------------------------------------------------------------------------ 		
%	FIGURE: Example Graph
%--------------------------------------------------------------------------
%---------------------------------------------------------------------------- 
%END FIGURE: Example Graph
%---------------------------------------------------------------------------
Figure \ref{Fig-Example-Metric} illustrates an example FD graph. Note that the directives are the leaf nodes. Initial Capability sets are formulated from the internal nodes $n_i, i=1, \ldots, 9$, as they represent functional abstractions, on the basis of their coupling and cohesion values. The next section defines these measures and describes their role in formulating Capabilities, the other activity of Phase I.
	
\subsection{Phase I: Formulation}
The objective of the formulation activity shown in Figure \ref{fig:FigNeedsToRequirements} is to establish initial sets of Capabilities from $G$. %from internal nodes of the FD graph $G$.
 An initial set is a meaningful combination of internal nodes and is termed as a \textit{slice}. There can be many slices from a single FD graph. We use cohesion $(Ch)$ and coupling $(Cp)$ measures to compute the overall cohesion-coupling value, $f(Ch,Cp)$, of each slice. This value %$(f(Ch,Cp))$ 
is used to determine the initial sets of Capabilities from all possible slices of $G$. In this section we first explain why we choose to measure cohesion and coupling of nodes. Then, we elaborate on each individual measure and its metric.
Finally, we discuss the construction of a slice and outline the process of selecting initial Capabilities based on $f(Ch,Cp)$. 

\subsubsection{Why cohesion \& coupling}
Capabilities are formulated so as to exhibit high cohesion and low coupling. Techniques of modularization suggest that these characteristics are typical of stable units \cite{Parnas1972} \cite{Yourdon1979}. Stability implies resistance to change; in the context of CE, we interpret stability as a property that accommodates change with minimum ripple effect. Ripple effect is the phenomenon of propagation of change from the affected source to its dependent constituents \cite{Haney1972}. Specifically, dependency links between modules behave as change propagation paths. The higher the number of links, the greater is the likelihood of ripple effect. Because coupling is a measure of interdependence between modules \cite{Stevens1974} we choose coupling as one indicator of stability of a module. In contrast, cohesion, the other characteristic of a stable structure, depicts the ``togetherness'' of elements within a module. Every element of a highly cohesive unit is directed toward achieving a single objective. 
We focus on maximizing functional cohesion, which indicates the highest level of cohesion \cite{Bieman1994} among all the other levels (coincidental, logical, temporal, procedural, communicational, and sequential) \cite{Yourdon1979} and therefore, is most desirable. In particular, a Capability has high functional cohesion if all its constituent elements, \textit{viz.} directives (later mapped to requirements), are devoted to realizing the function represented by the Capability. As a general observation as the cohesion of a unit increases the coupling between the units decreases. However, this correlation is not exact \cite{Yourdon1979}. Therefore, we develop specific metrics to measure the coupling and cohesion values of internal nodes in $G$, and thereby, formulate initial sets of Capabilities. 

\subsubsection{Cohesion Measure}
A unit has functional cohesion if it focuses on executing exactly one basic function. Yourdon and Constantine \cite{Yourdon1979} state that every element in a module exhibiting functional cohesion ``is an integral part of, and is essential to, the performance of a single function''. By the virtue of construction, in the FD graph the function of each child node is essential to achieving the function of its immediate parent node. Note that, neither the root nor the leaves of an FD graph can be considered as a Capability. This is because the root indicates the mission of the system, which is too holistic, and the leaves symbolize directives, which are too reductionistic in nature. Both of these entities lie on either extreme of the abstraction scale, and thereby, conflict with the objective of avoiding such polarity when developing complex emergent systems \cite{Heylighen1989}. Thus, only the internal nodes of an FD graph are considered as potential Capabilities. In addition, these internal nodes depict functionalities at different levels of abstraction, and thereby, provide a representative sample for formulating Capabilities. We develop the cohesion measure for internal nodes by first considering nodes whose children are only leaves. We then generalize this measure for any internal node in the graph. 

\paragraph{Measure for internal nodes with only leaves as children}
Internal nodes with only leaves as children represent potential Capabilities that are linked directly to a set of directives. In Figure \ref{Fig-Example-Metric} these are nodes $n_5, n_6, n_7, n_8, n_9$. Directives are necessary to convey and develop an in depth understanding of the system functionality and yet, by themselves, lack sufficient detail to dictate system development. Failure to implement a directive can affect the functionality of the associated Capability with varying degrees of impact. We hypothesize that the degree of impact is directly proportional to the relevance of the directive to the functionality. Consequently, the greater the impact, the more essential the directive. This signifies  the strength of relevance of a directive and is symptomatic of the associated Capability's cohesion. Hence, the  \textit{relevance} of a directive to the functionality of a unit is an indicator of the unit's cohesion.

The failure to implement a directive can be interpreted as a risk. Therefore, we use existing risk impact categories: Catastrophic, Critical, Marginal and Negligible \cite{Boehm1989} to guide the assignment of relevance values. 
Each impact category is well-defined and has an associated description. This  is used to estimate the relevance of a directive on the basis of its potential impact. For example, in Table \ref{tab:ScaleOfRelevance} negligible impact is described to be only an inconvenience, whereas a catastrophic impact implies complete failure. This signifies that the relevance of a directive with negligible impact is much lower when compared to a directive with catastrophic impact.
Intuitively, the impact categories are ordinal in nature. However, we conjecture that the associated relevance values are more than merely ordinal. The issue of determining the natural measurement scales \cite{Stevens1946} of cohesion and other software metrics is an open problem \cite{Briand1996}. Therefore, we refrain from subscribing both, the attribute in question \textit{i.e.} cohesion and its metric \textit{i.e.} function of relevance values, to a particular measurement scale. 
Rather than limiting ourselves to permitted analysis methods as defined by Stevens \cite{Stevens1946} we let the objective of our measurement --- computing the cohesion of a node to reflect the relevance of its directives --- determine the appropriate statistic to be used \cite{Velleman1993}. 

We assign values to indicate the relevance of a directive based on the perceived significance of each impact category; these values are normalized to the [0,1] scale. 
The categories and their associated relevance values are listed in Table \ref{tab:ScaleOfRelevance}. We estimate the \textit{cohesion} of an internal node as the average of the relevance values of all its directives. The \textit{arithmetic mean} is used to compute this average as it can be influenced by extreme values. This thereby captures the importance of directives with catastrophic impact or the triviality of directives with negligible impact, and affects the resulting average appropriately, to reflect the same. 
%------------------------------------------------------------------------ 			
%TABLE: Scale of Relevance
%--------------------------------------------------------------------------
\begin{table}[h]
%\caption{\em Relevance Values}
\centering
\begin{tabular}{|l|l|c|}
\hline
\textbf{\textsc{Impact}} & \textbf{\textsc{Description}} & \textbf{\textsc{Relevance}}  \\
\hline \hline
%Catastrophic &  Task failure & 0.71-1.00\\
Catastrophic &  Task failure & 1.00\\
\hline
%Critical & Task success questionable & 0.31-0.70\\
Critical & Task success questionable & 0.70\\
\hline
%Marginal &  Reduction in technical performance & 0.11-0.30\\
Marginal &  Reduction in technical performance & 0.30\\
\hline
%Negligible& Inconvenience/ nonoperational impact  & 0.00-0.10\\
Negligible& Inconvenience/ nonoperational impact  & 0.10\\
\hline
%Null& No impact  & 0.00\\
%\hline
%\caption{\em Impact Categories with associated relevance}
\end{tabular}
\setlength{\abovecaptionskip}{-0.35cm}   % 0.5cm as an example
\setlength{\belowcaptionskip}{0cm}   % 0.5cm as an example
\caption{\em Relevance Values}
\label{tab:ScaleOfRelevance}

\end{table}
%------------------------------------------------------------------------ 			
%END TABLE: Scale of Relevance
%--------------------------------------------------------------------------
Every parent-leaf edge is associated with a relevance value $Rel(v,n)$ indicating the contribution of directive $v$ to the cohesion of parent node $n$. For example in Figure \ref{Fig-Example-Metric}, $Rel(d_1,n_5)=0.3$. Note that, we measure the relevance of a directive only to its immediate functionality. For an FD graph $G=(V,E)$ we denote relevance of a directive $d$ to its parent node $n$ as $Rel(d,n)$ where $d,v \in V$, $(n,d)\in E$, $outdegree(d)=0$ and $outdegree(n)>0$. Formally, the cohesion measure of a potential Capability that is directly associated with a set of directives \textit{i.e.} the cohesion measure of an internal node $n \in V$ with $t$  leaves as its children ($t>0$), is given by computing the arithmetic mean of relevance values:

\begin{center}
$Ch(n) =\; \dfrac { \overset{t} {\underset {\scriptstyle i=1} \sum} Rel(d_i,n)}{t}$ \; 
\end{center}

For example in Figure \ref{Fig-Example-Metric}, $Ch(n_7)=0.525$. The cohesion value ranges between $0$ and $1$. A Capability with a maximum cohesion of $1$ indicates that every constituent directive is of the highest relevance. 
\paragraph{Measure for internal nodes with only non-leaf children}
Cohesion measure for internal nodes with only non-leaf children is computed differently. This is because the relevance value of a directive is valid only for its immediate parent and not for its ancestors. For example, the functionality of node $n_1$ in Figure \ref{Fig-Example-Metric} is decomposed into nodes $n_5$ and $n_6$. This implies that the functionality of $n_1$ is directly dependent on the attainment of the functionality of both $n_5$ and $n_6$. Note that $n_1$ has only an indirect relationship to the directives of the system. In addition, the degree of influence that $n_5$ and $n_6$ each have on parent $n_1$ is influenced by their size (number of constituent directives). Therefore, the cohesion of nodes that are \textit{parents with non-leaf children} is a weighted average of the cohesion of their children. Here, the weight is the size of a child node in terms of its constituent directives. This indicates the child's contribution towards the parent's overall cohesion. 
The rationale behind this is explained by the definition of cohesion, which states that a node is highly cohesive if every constituent element is focused on the same objective, \textit{i.e.} the node's functionality.

Formally, the cohesion measure of an internal node $n$ with $t>1$ non-leaf children is:
\begin{displaymath}
Ch(n) =	 \dfrac{\overset{t}{\underset {\scriptstyle i=1} \sum} (size(v_i). 					Ch(v_i))}{\overset{t} {\underset {\scriptstyle i=1} \sum} size(v_i)}	
\end{displaymath}

such that $(n,v_i)\in E $  and,

\begin{displaymath}
				size(n)= \begin{cases}
	
								\overset{t}{\underset {\scriptstyle i=1} \sum } size(v_i) & (n,v_i) \in E; outdegree(v_i)>0;\\
								 1 & outdegree(n)=0 
									\end{cases}
					\end{displaymath}
 
In the case where $outdegree(n)=1$, \textit{i.e.} the node has only one child $v$ (say), then the $Ch(n)$ is the $Ch(v)$; if $outdegree(n)=0$, \textit{i.e.} $n$ is a leaf (directive), $Ch(n)$ is not applicable. 

\subsubsection{Coupling Measure}
As with cohesion, the concept of coupling was introduced by Stevens \textit{et al.} \cite{Stevens1974} as the ``measure of the strength of association established by a connection from one module to another''. Coupling is also characterized as the degree of interdependence between modules. 
The objective of CE is to identify minimally coupled nodes as initial Capabilities. 
A Capability is related to another Capability only through its constituent directives, \textit{i.e.} the coupling between Capabilities is the measure of the dependencies between their respective directives. Thus, we first discuss the coupling between directives, and then develop the coupling measure for Capabilities. 

\paragraph{\textit{Coupling between directives}}
Generally, metrics that measure coupling between modules utilize data from the source code or the system design. However, to measure the coupling between directives we have neither design information nor implementation details at our disposal. We only have the structural information provided by the FD graph. 
In particular, we consider an undirected version $G'$ (shown in Figure \ref{Fig-Example-Metric-Coupling}) of the FD graph $G$ where $G'=(V,E')$ and $E'$ is the set of undirected edges. 
We denote coupling between directives (leaf nodes) $d_x$ and $d_y$ as $Cp(d_x,d_y)$. Note that $Cp(d_x,d_y) \neq Cp(d_y,d_x) $ as $Cp(d_x,d_y)$ is the dependency of $d_x$ on $d_y$, which can be quantified by measuring the effect on $d_x$ when $d_y$ changes. Similarly, $Cp(d_y,d_x)$ indicates the dependency of $d_y$ on $d_x$. In general, we hypothesize coupling as a function of two components: distance and probability of change.

\begin{itemize}
	\item \textit{Distance:} 
We know that directives associated with the same Capability are highly functionally related. In $G'$, this is represented by leaves that share the same parent node. However, relatedness between directives decreases with increasing distance between them. We define distance between directives $u, v \in V$ as the number of edges in the shortest undirected path between them and denote it as $dist(u,v)$. By choosing the shortest path we account for the worst case scenario of change propagation. Specifically, the shorter the distance, the greater the likelihood of impact due to change propagation.

In Figure \ref{Fig-Example-Metric-Coupling}, $d_1$ and $d_2$ are directives of the same parent $n_5$ and so are highly related with $dist(d_1,d_2)=2$. In contrast, $d_1$ and $d_9$ have a lower relatedness with $dist(d_1,d_9)=6$  as they are connected only through common ancestors. % $n_5, n_1, n_6, n_2, n_7$. 
The shortest paths connecting $d_1$ and $d_2$ and, $d_1$ and $d_9$ are highlighted in Figure \ref{Fig-Example-Metric-Coupling}. Thus, from the distance measure we conclude that $d_1$ is less likely to be affected by a change in $d_9$ than a change in $d_2$. Consequently, $Cp(d_1,d_9) < Cp(d_1,d_2)$. Hence, for any two directives $u$ and $v$ we deduce:
 
 \begin{displaymath}
  Cp(u,v) \propto \dfrac{1}{dist(u,v)}\\
               %\hspace{1.3in} $dist(u,v)$
  \end{displaymath}    
%------------------------------------------------------------------------ 		
%	FIGURE: Example Graph
%--------------------------------------------------------------------------
\begin{figure}[htbp]
\centering
% trim = left bottom right top
\includegraphics[trim = 5mm 170mm 0mm 15mm, clip, width=9.5cm]{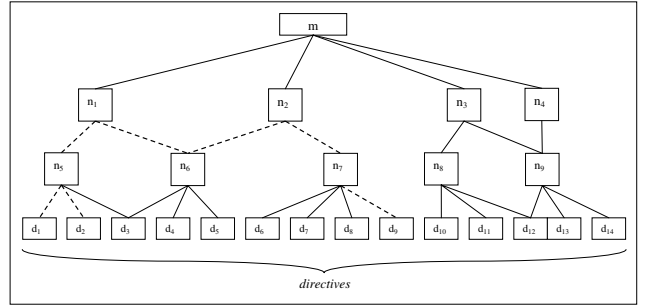}
\caption{\em Undirected FD Graph $G'=(V,E')$}
\label{Fig-Example-Metric-Coupling}
\end{figure}
%---------------------------------------------------------------------------- 
%END FIGURE: Example Graph
%---------------------------------------------------------------------------
\item \textit{Probability of Change:} 
We are interested in choosing internal nodes that are minimally coupled as initial Capabilities. Minimal interconnections reduce the likelihood of a ripple effect phenomenon. We know that coupling between Capabilities is a function of coupling between their respective directives. As mentioned earlier, if $u$ and $v$ be directives then $Cp(u,v)$ can be quantified by measuring the effect on $u$ when $v$ changes. However, we still need to compute the probability of occurrence of such a ripple effect. This implies computing the probability that a directive might change. Therefore, $Cp(u,v)$ also needs to factor in the probability of directive $v$ changing: $P(v)$.
Consequently, the coupling between two directives $u$ and $v$ is computed as:
\begin{displaymath}
  Cp(u,v) = \dfrac{P(v)}{dist(u,v)} \\
  \end{displaymath}

This metric signifies the coupling between directives $u$ and $v$ as the probability that a change in $v$ propagates through the shortest path and affects $u$. 
\end{itemize}

\paragraph{\textit{Coupling between Capabilities}} Capability  $p$ is coupled with Capability $q$  if a change in $q$ affects $p$. Note that $Cp(p,q)$ is the measure that $p$ is coupled with $q$ and so, $Cp(p,q) \neq Cp(q,p)$. In particular, a change in $q$ implies a change in one or more of its constituent directives. Therefore, the coupling measure for Capabilities is determined by the coupling between their respective directives. However, it is possible that $p$ and $q$ share common directives. In such a case, we need to make a decision about the membership of these directives and ensure that they belong to exactly one Capability. This is reflective of the actual system, where any functionality is implemented only once and is not duplicated. Criteria such as the relevance value of the directive or its contribution to the overall cohesion may be used to resolve this issue. For now, we use the former criteria. 

In terms of $G$ we define the set of leaves (directives) associated with an internal node $n$ as:
\begin{center}
$D_{n}=\{x|\hspace{0.15cm} \exists path(n,x); outdegree(x)=0; n,x \in V \}$
\end{center} 
where
$path(n,x)$ is a set of directed edges connecting $n$ and $x$. %Also, $\underset {\scriptstyle i}\cap D_i = \phi, 1<i< size(root)$. 
For example the set of leaves associated with the internal node $n_3 \in V$ is $D_{n_3}=\{d_{10}, d_{11}, d_{12}, d_{13}, d_{14}\}$. Now consider $Cp(n_5,n_6)$, from Figure \ref{Fig-Example-Metric-Coupling}, which is the coupling between internal nodes $n_5$ and $n_6$. As $Cp(n_5,n_6)$ quantifies the effect on $n_5$ when $n_6$ changes \textit{i.e.} we need to compute the effect on the directives associated with $n_5$: $d_1, d_2, d_3$ when the directives associated with $n_6$: $d_4, d_5$ change. We compute coupling by associating the common directive $d_3$ with $n_5$ and not $n_6$ because $Rel(d_3,n_5) > Rel(d_3,n_6)$. We use the relevance value to decide the membership of a directive. Therefore, $D_{n_5}=\{d_1,d_2,d_3\}$ and $D_{n_6}=\{d_4,d_5\}$. The coupling between $n_5$ and $n_6$ is given by:

\begin{center}
$Cp(n_5,n_6)= \dfrac{\displaystyle \sum_{d_i \in D_{n_5}} 
\displaystyle  \sum_{d_j \in D_{n_6}} Cp(d_i,d_j)} {|D_{n_5}|. |D_{n_6}|} $
\end{center}

where $|D_{n_5}|$ is the cardinality of $D_{n_5}$. \\

Generalizing, the coupling measure between any two internal nodes $p, q \in V$, %for graph $G'$ 
where $outdegree(p)>1, outdegree(q)>1$ and $D_p \cap D_q=\phi$ is:

\begin{center}
$Cp(p,q)= \dfrac{\displaystyle \sum_{d_i \in D_{p}} 
\displaystyle  \sum_{d_j \in D_{q}} Cp(d_i,d_j)} {|D_{p}|. |D_{q}|} $
\end{center}

\bigskip
where $Cp(d_i,d_j)= \dfrac{P(d_j)}{dist(d_i,d_j)}$ and $P(d_j)= \dfrac{1}{|D_q|}$. \\

$P(d_j)$ is the probability that directive $d_j$ changes among all other directives associated with the node $q$. 

\subsubsection{ Initial Capabilities Sets}
The cohesion and coupling measures are used to formulate initial sets of Capabilities from the FD graph. However, prior to the application of these measures, we determine what combinations of internal nodes  are meaningful enough to be considered as Capabilities. For example, in FD graph $G$ of Figure \ref{Fig-Example-Metric} the set $\{n_1,n_5,n_6\}$ is an unsound combination of Capabilities as they are a redundant portrayal of only a part of the system functionality. Recall that Capabilities are functional abstractions that form the foundation of a complex emergent system, and thereby, need to be formulated with sound principles and rules. 

We first identify valid combinations of internal nodes termed $slices$ from an FD graph. Then, we apply the measures of coupling and cohesion on these slices to determine the initial sets of Capabilities. Note that each node of a slice is a potential Capability. 
For an FD graph $G=(V,E)$ we define slice $S$ as a subset of $V$ % $S \subset V$ 
where the following constraints are satisfied:

\begin{enumerate}
\item \textit{Complete Coverage of Directives:} We know that a Capability is associated with a set of directives, which are finally mapped to system requirement specifications (see Figure \ref{fig:FigNeedsToRequirements}). Consequently, a set of initial Capabilities of the system has to encompass all the directives resolved from user needs. The leaves of the FD graph constitute the set of all directives in a system. We ensure that each directive is accounted for by some Capability, by enforcing the constraint of complete coverage given by 
${\overset{m}{\underset{\scriptstyle i=1} \bigcup} D_i}=\{L\}$, where
\begin{itemize}
	\item $D_i$ denotes the set of leaves associated with the $i^{th}$ node of slice $S$
	\item	$L=\{u \in V| outdegree(u)=0\}$ denotes the set of all leaves of $G$
	\item	$m=|S|$ 
\end{itemize}

\item \textit{Unique Membership for Directives:} In the context of directives, by ensuring that each directive is uniquely associated with exactly one Capability we avoid implementing redundant functionality. Otherwise, the purpose of using slices to determine Capabilities as unique functional abstractions is defeated. We ensure the unique membership of directives by the constraint ${\overset{m}{\underset{\scriptstyle i=1} \bigcap} D_i}=\{\phi\}$. 

\item \textit{System Mission is not a Capability:} The root is the high level mission of the system and cannot be considered as a Capability. The  cardinality of a slice containing the root can only be one.
This is because including other nodes with the root in the same slice violates the second constraint. 
Hence, $ \{\forall u \in S, indegree(u) \neq 0 \}$. 

\item \textit{Directive is not a Capability:} A leaf represents a directive, which is a system characteristic. A slice that includes a leaf fails to define the system in terms of its functionality and focuses on describing low level details. 
Hence, $\{\forall u \in S, outdegree(u) \neq 0 \}$.
\end{enumerate}

For example $S = \{n_1, n_7, n_3\}$ is a valid slice for the graph illustrated in Figure \ref{Fig-Example-Metric}. Note that criteria such as the relevance value of a directive or its contribution to the associated node's cohesion value are used to decide the membership of a directive so that it is unique, satisfying the second constraint. Let there be $p > 1$ number of slices computed from graph $G$. We use the previously defined measures to rank the slices based on their values of coupling and cohesion. Based on this ranking we determine initial sets of Capabilities.

Let $Ch_i$ and $Cp_i$ denote the cohesion and coupling values of the $i^{th}$ node of slice $S_j$ respectively, where $S_j=\{n_i, 1\leq i \leq q, q=|S_j|\}, 1 \leq j \leq p$. We compute $f_j(Ch,Cp)$, a function of the cohesion and coupling values of all nodes in $S_j$ to represent the overall cohesion-coupling value of the slice. We rank the $p$ slices based on their cohesion-coupling value $f(Ch,Cp)$ and choose those slices with an above average value as initial sets of Capabilities. 

The initial sets of Capabilities, described above, form the output of Phase I. The next phase of the CE process as shown in Figure \ref{fig:FigNeedsToRequirements} is Phase II. In Phase II, we apply an MDO approach on the initial sets to determine the most optimal set of Capabilities. The optimized Capabilities and their associated directives are then mapped to system requirements. Thus, \textit{optimization} and \textit{mapping} are the two major activities of Phase II. 

\subsection{Phase II: Optimization}
\label{sec:optimization}
The initial sets of Capabilities are in essence, combinations of internal nodes selected from the FD graph. These Capabilities and their directives possess valuable domain knowledge and represent user needs. However, they are too coarse and unrefined to dictate system development in the solution domain. Hence, we need to optimize these Capabilities with respect to constraints germane to complex emergent system development. In particular, we focus on three %optimize the initial sets of Capabilities with respect to three 
specific constraints: overall cohesion-coupling value $f(Ch,Cp)$, technology feasibility $tf$ and schedule $sched(order, time)$. The initial set whose values of $f(Ch,Cp)$, $tf$ and $sched(order,time)$ are optimal when compared to all other sets of Capabilities is selected using the MDO approach. Conceptually, we desire to maximize the objective function $z$ subject to the previously defined constraints. This is described as: 

\textbf{Objective Function:} 
\begin{center}
\textit{Maximize} $z(f(Ch,Cp),tf,sched(order,time))$
\end{center}

\textbf{Constraints:}

\begin{center}

$tf \geq$ \textsc{\small{$tf_{MIN}$}}

$sched \leq$ \textsc{\small{$sched_{MAX}$}}	

$f(Ch,Cp)\geq$ \textsc{\small{$f_{MIN}(Ch,Cp)$}}

\end{center}

%------------------------------------------------------------------------ 			
%FIGURE: MDO
%--------------------------------------------------------------------------
\begin{figure}[ht]
\centering
% trim = left bottom right top

\includegraphics[trim = 5mm 145mm 0mm 0mm, clip, width=11 cm]{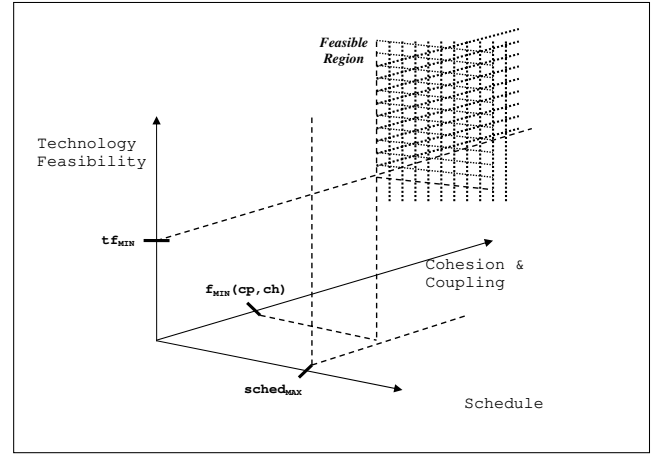}
\caption{\em Feasible Region}
\label{fig:Fig-MDO}
\end{figure}
%---------------------------------------------------------------------------- 		
%END FIGURE: MDO
%-----------------------------------------------------------------------------
\noindent
Note that values $tf_{MIN}$, $sched_{MAX}$, $f_{MIN}(Ch,Cp)$ can be defined by the user. Figure \ref{fig:Fig-MDO} illustrates this conceptually, depicting the feasible region. %Exploring the MDO approach is part of our current ongoing research.
Since we have already discussed $f(Ch,Cp)$, we now explain the other two constraints: $tf$ relating to technology advancement and $sched(order,time)$ derived from the implementation schedule. 

\paragraph{Technology Advancement}

We examine two possible scenarios, caused by technology advancement, when incorporating technology in a system: 
\begin{itemize}
	\item \textit{Technology Obsolescence:} 
Given the rapid rate of hardware advancements, a lengthy development period of a complex emergent system can render the initial technology requirements  invalid. Consequently, the technology of an existing Capability becomes obsolete.
		
	\item \textit{Technology Infusion:} 
The functionality expected of a system may undergo substantial modification over a long period of time requiring the introduction of new Capabilities. This results in the infusion of new technology into the existing system.
\end{itemize}

\noindent
Intuitively, we know that by minimizing the coupling between Capabilities, the impact of change relative to technology advancement is reduced. In addition, we hypothesize that cohesion also plays a vital role. In Capabilities with high cohesion every element is highly focused on a single function. Consequently, elements of a functionally cohesive Capability are strongly tied to the underlying technology, as this technology assists in implementing the functionality. Hence, replacing technology of an existing Capability is easier when it is highly cohesive. We use the term $tf$, \textit{i.e.} technology feasibility, to indicate the feasibility of currently available technology to implement an initial set of Capabilities. More specifically, $tf_S^i$ is the feasibility of the available technology to satisfactorily develop slice $S$ at the time instant $i$.

\paragraph{Schedule}
Similar to technology feasibility we also consider the implementation schedule as being a constraint on selecting slices. We theorize that schedule is a function of order and time; $sched(order, time)$. \textit{Order} is the sequence in which the Capabilities of a slice need to be developed. In particular, it is likely that certain functionalities have a higher priority of development than others. Hence, the order of developing functionalities is crucial in the selection of Capabilities. Furthermore, some functionalities may have to be implemented within a specific time period. Thus, $time$ is also a factor in determining the schedule. In conclusion, when selecting slices we focus on the constraints of coupling-cohesion, technology feasibility and schedule, to combat the factors of change \textit{viz.} volatility and technology advancement. 

\subsection{Phase II: Mapping to Requirements}
The final activity of the CE process, as shown in Figure \ref{fig:FigNeedsToRequirements}, is the mapping of directives to system requirements. We claim that there is a one-many mapping from a directive to a requirement. Both entities are defined at a reductionistic level of abstraction and share the objective of signifying the characteristics of a system. Therefore, we hypothesize that the process of mapping is uncomplicated.

\section{Validation}
In this section we describe our ongoing research activities to validate the efficacy of the CE approach for constructing change-tolerant systems. In general the best approach to asserting the validity of CE  is to employ a longitudinal study spanning the development of a complex emergent system. 
However, such an approach warrants a lengthy time-period. Alternatively, we choose to validate our theory on an existing system that exhibits the characteristics of a complex emergent system and possesses a change history archive. The following sections describe this system, examine its appropriateness for validation purposes, outline the validation procedures and discuss some preliminary observations.

\subsection{System Characteristics}
The system being used for validation purposes is Sakai Collaboration and Learning Environment, an open-source, enterprise-scale software application. It is being developed by an international alliance of several universities spanning four continents \cite{Whyte2006}. The current Sakai system consists of about 80,000 lines of code and its complexity is futher compounded by distributed development.
 The high-level mission of this system is to realize specific research, learning, collaboration and academic needs of universities and colleges. The system is constantly evolving to accommodate the needs of its 300,000+ users. System increments that incorporate new functionalities are released on a yearly basis. Also, the overall system is envisioned to be used for an extended time-period. Hence, the Sakai system exhibits characteristics of complex emergent systems and appears suitable for the purpose of validating our CE approach. 

\subsection{Outline of the Validation Approach}
We have constructed the FD graph, $G_{sk}$, for the Sakai system based on extensive documentation of user needs, long-term feature requests and results of community polls. The graph has also been validated by the members of the Sakai project to ensure that it is a true decomposition of user needs. 
We have computed 85 valid slices from a possible $1152921504606846976$  
combinations of nodes. Our hypothesis is that an optimal slice appropriately identified by CE, say $S_{ce}$, is more change-tolerant than the slice implemented in the actual Sakai system, $S_{sk}$. Both $S_{ce}$ and $S_{sk}$ are among the set of valid slices previously determined from $G_{sk}$.
To test the hypothesis we examine the ripple-effects of change in both the slices. The comprehensive change history maintained in Sakai archives facilitate the ripple-effect analysis in $S_{sk}$. We inspect the code to trace the effect of similar changes in $S_{ce}$. Several different scenarios of change --- modified requirements, deleted needs, addition of new features --- are to be analyzed. Presently, we quantify the impact of change as the number of affected entities. %($\#AEnt$). 
These entities can be requirements, implementation modules, system functionalities and so on.  

\subsection{Preliminary Empirical Observations}
A preliminary analysis of $G_{sk}$ has resulted in several informative observations related to the construction of change-tolerant systems. We outline four of them below:

\subsubsection{Common functionalities}
The graph structure indicates that there is a relationship between the number of intersection edges of a node and its coupling measure with other nodes. Recall that an intersection edge indicates common functionalities.
In particular, the higher the number of intersection edges from an internal node, the greater is its coupling value. We observe that the addition of an intersection edge might provide a shorter path of traversal between directives, and thereby, result in increased coupling. To what extent, if any, can this observation be used in guiding the design of complex emergent systems?

\subsubsection{Factoring in Level of Abstraction}
In certain cases, we observe that an optimal slice of the FD graph consists of nodes defined at the highest level of abstraction. This is because our cohesion and coupling measures are averages derived from bottom-up computations, and thereby, tend to identify nodes closest to the root as being optimal.
In general, there is a relation between the abstraction level and the size (number of associated directives) of a node. This is exemplified by an FD graph that is a complete tree, where nodes at a lower level of abstraction are of a smaller size. From a software engineering perspective, it is prudent that the Capabilities of a system are not only  highly cohesive and minimally coupled, but are also of reduced size. Hence, nodes of a lower abstraction that are marginally more coupled but are of a smaller size, are better suited for system development. Therefore, the abstraction level, ascertained from a top-down decomposition of the graph, should be utilized in conjunction with the bottom-up measures of cohesion and coupling to determine optimal slices.
 
\subsubsection{Coupling-Cohesion trend}
The computation of cohesion and coupling metrics is independent of each other. 
However, we observe that on an average, in a slice, nodes \textit{viz.} Capabilities that have high cohesion values also exhibit low coupling with other nodes. This is certainly in line with with desirable software engineering characteristics.

\subsubsection{Schedule}
Coupling between two nodes is determined in part by the sizes of each node. Therefore, the coupling between two nodes can be asymmetric. For example, in Figure \ref{Fig-Example-Metric-Coupling},
$n_1$ and $n_9$ are of different sizes and so, $Cp(n_1,n_9) \neq Cp(n_9,n_1)$. Consequently, the coupling measure can assist in choosing an implementation order of Capabilities that potentially minimizes the impact of change. Note that permuting a slice of nodes produces different sequences of implementation, each of whose coupling value can be computed. This observation implies that the coupling measure can help define the criteria for determining an implementation schedule, which is a function of order and time as discussed in Section
\ref{sec:optimization}.

\section{Conclusion}
Complex emergent systems need to be change-tolerant, as they have lengthy development cycles. Requirements and technology often evolve during such development periods, and thereby, inhibit a comprehensive up-front solution specification. Failing to accommodate changed requirements or to incorporate latest technology results in an unsatisfactory system, and thereby, invalidates the huge investments of time and money. Recent history of system failures provides ample evidence to support this fact. 
We propose an alternative approach of development termed CE to develop change-tolerant systems. It is a scientific, disciplined and deliberate process for defining Capabilities as functional abstractions, which are the building blocks of the system. Capabilities are designed to exhibit high cohesion and low coupling, which are also desirable from a software engineering perspective, to promote change-tolerance. Also, the CE process touts an MDO approach for selecting an optimal set of Capabilities that accommodates the constraints of technology advancement and development schedule. CE is a recursive process of selection, optimization, reorganization and hence, the stabilization of Capabilities. We envision that the Capabilities based approach provides a high level development framework, for  complex emergent systems, accommodating change and facilitating evolution with minimal impact. 
\bibliography{HICCS}
% that's all folks
\end{document}